\documentclass[12pt]{article}
\usepackage{amsfonts,amssymb,amsbsy,epsf,rotate}
\def\slashchar#1{\setbox0=\hbox{$#1$}           
   \dimen0=\wd0                                 
   \setbox1=\hbox{/} \dimen1=\wd1               
   \ifdim\dimen0>\dimen1                        
      \rlap{\hbox to \dimen0{\hfil/\hfil}}      
      #1                                        
   \else                                        
      \rlap{\hbox to \dimen1{\hfil$#1$\hfil}}   
      /                                         
   \fi}                                         %
\begin{document}
\title{{\large\sc Possibility of Measuring the {\bf CP}-Violation
$\boldsymbol \gamma$-parameter
in Decays of $\boldsymbol \Xi_{bc}$ Baryons}}
\author{V.V.Kiselev, O.P.Yushchenko\\[2mm]
\small Russian State Research Center ``Institute for
High Energy Physics'', \\
\small  Protvino, Moscow Region, 142281 Russia\\
\small E-mail: kiselev@th1.ihep.su, Fax: 7-0967-744937}
\date{}

\maketitle
\begin{abstract}
  The model-independent method based on the triangle ideology is
  implemented to extract the CKM-matrix angle $\gamma$ in the decays
  of doubly heavy baryons containing the charmed and beauty quarks. We
  analyze a color structure of diagrams and conditions for
  reconstructing two reference-triangles by tagging the flavor and CP
  eigenstates of $D^0\leftrightarrow \bar D^0$ mesons in the fixed
  exclusive channels. The characteristic branching ratios are
  evaluated in the framework of a potential model setting a parametric
  dependence on the hadronic matrix elements for the decay rates.
\end{abstract}

\section{Introduction}
\label{sec:1}

The current success in the experimental study of decays with the
CP-violation in the gold-plated mode of neutral $B$-meson by the
BaBar and Belle collaborations \cite{BB} allows one to extract the
CKM-matrix angle $\beta$ in the unitarity triangle by the
model-independent method.  The intensive efforts are intended in
the physical programs on the $B$ and $B_s$ mesons at the hadron
colliders both the active Tevatron \cite{BRunII} and prospective
LHC. Due to the relatively high cross-sections the doubly heavy
hadrons such as the $B_c$ meson and baryons $\Xi_{bc}$,
$\Omega_{bc}$ and $\Xi_{cc}$, $\Omega_{bc}$ would be copiously
produced at such the machines \cite{revbc,revqq}. In addition to
the indirect or model-dependent measurements of unitarity triangle
\cite{Buras}, there is an intriguing opportunity to extract the
angle $\gamma$ in the model-independent way using the strategy of
reference triangles \cite{Gronau} in the decays of doubly heavy
hadrons. This ideology for the study of CP-violation in $B_c$
decays was originally offered by M.Masetti \cite{Masetti} and
investigated by R.Fleischer and D.Wyler \cite{FW}. In this letter
we extend the method to study the decays of doubly heavy baryons
containing the charmed and beauty quarks.

\begin{figure}[th]
  \begin{center}
\setlength{\unitlength}{1mm}
\begin{picture}(120,25)
\put(-15,0){\epsfxsize=55\unitlength \epsfbox{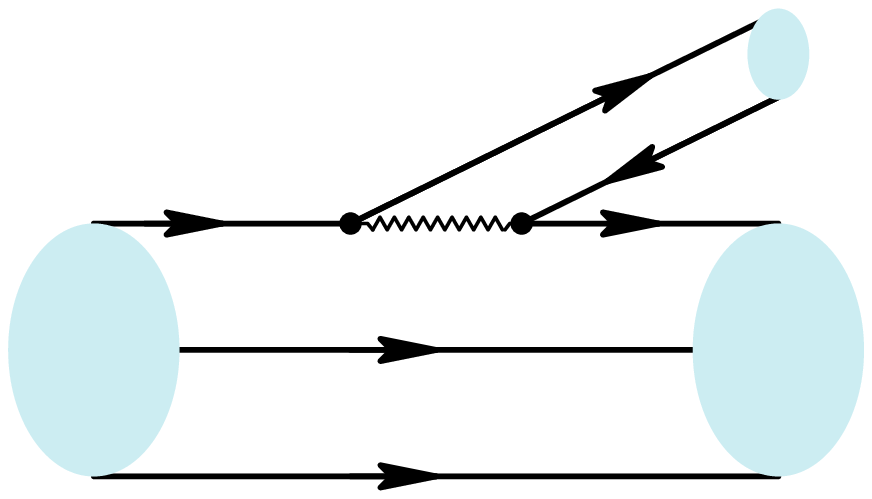}}
\put(33,0){\epsfxsize=55\unitlength \epsfbox{n1.eps}}
\put(85,0){\epsfxsize=55\unitlength \epsfbox{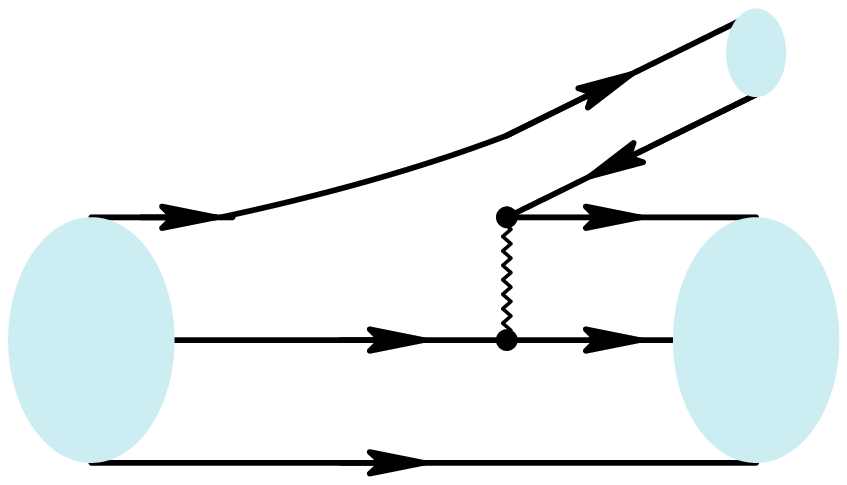}}
\put(-5,7.5){$\Xi^0_{bc}$}
\put(29.5,7.5){$\Xi^0_{c}$}
\put(33,22){$\bar D^0$}
\put(43,7.5){$\Xi^0_{bc}$}
\put(77.5,7.5){$\Xi^0_{c}$}
\put(81,22){$D^0$}
\put(95,7.5){$\Xi^0_{bc}$}
\put(129.5,7.5){$\Xi^0_{c}$}
\put(133,22){$D^0$}
\put(1.5,16.5){$\scriptstyle b$}
\put(12.5,10.5){$\scriptstyle c$}
\put(12.5,3.7){$\scriptstyle d$}
\put(22.5,23){$\scriptstyle u$}
\put(23,18.8){$\scriptstyle c$}
\put(26,15.7){$\scriptstyle s$}
\put(49.5,16.5){$\scriptstyle b$}
\put(60.5,10.5){$\scriptstyle c$}
\put(60.5,3.7){$\scriptstyle d$}
\put(70.5,23){$\scriptstyle c$}
\put(71,18.8){$\scriptstyle u$}
\put(74,15.7){$\scriptstyle s$}
\put(101.5,16.5){$\scriptstyle c$}
\put(112.5,10.5){$\scriptstyle b$}
\put(112.5,3.7){$\scriptstyle d$}
\put(122.5,23){$\scriptstyle c$}
\put(123,18.8){$\scriptstyle u$}
\put(126,15.7){$\scriptstyle s$}
\put(123.5,10.5){$\scriptstyle c$}
\end{picture}
    \caption{The diagrams of $b$-quark decay contributing to the weak
     transitions $\Xi^0_{bc}\to D^0\Xi^0_c$ and
     $\Xi^0_{bc}\to \bar D^0\Xi^0_c.$}
    \label{fig:1}
  \end{center}
\end{figure}

To begin, we mention the necessary conditions for extracting the
CP-violation effects in the model-independent way.
\begin{enumerate}
\item
Interference. The measured quantities have to involve the
amplitudes including both the CP-odd and CP-even phases.
\item
  Exclusive channels. The hadronic final state has to be fixed in
  order to isolate the definite matrix elements of CKM matrix, which can exclude the
  interference of two CP-odd phases with indefinite CP-even phases due
  to strong interactions at both levels of the quark structure and the
  interactions in the final state.
\item Oscillations. The definite involvement of the CP-even phase is
  ensured by the oscillations taking place in the systems of
  neutral $B$ or $D$ mesons, wherein the CP-breaking effects can be
  systematically implemented.
\item Tagging. Once the oscillations are involved, the tagging of both
  the flavor and $CP$ eigenstates is necessary for the complete
  procedure.
\end{enumerate}
The gold-plated modes in the decays of neutral $B$ mesons involve
the oscillations of mesons themselves and, hence, they require the
time-dependent measurements. In contrast, the decays of doubly
heavy hadrons such as the $B_c$ meson and $\Xi_{bc}$ baryons with
the neutral $D^0$ or $\bar D^0$ meson in the final state do not
require the time-dependent measurements. The triangle ideology is
based on the direct determination of absolute values for the set
of six decays: the decays of baryon in the tagged $D^0$ meson, the
tagged $\bar D^0$ meson, the tagged CP-even state, and those of
the anti-baryon. To illustrate, let us consider the decays of
$$
\Xi^0_{bc}\to D^0\Xi^0_c, \quad\mbox{and}\quad
\Xi^0_{bc}\to \bar D^0\Xi^0_c.
$$
The corresponding diagrams with the decay of $b$-quark are shown
in Fig. \ref{fig:1}. We stress that two diagrams of the baryon
decay to $D^0$ has the additional negative sign caused by the
Pauli interference of two charmed quarks, while the color factors
is analyzed in the next section.

\begin{figure}[th]
  \begin{center}
\setlength{\unitlength}{1mm}
\begin{picture}(120,25)
\put(-15,0){\epsfxsize=55\unitlength \epsfbox{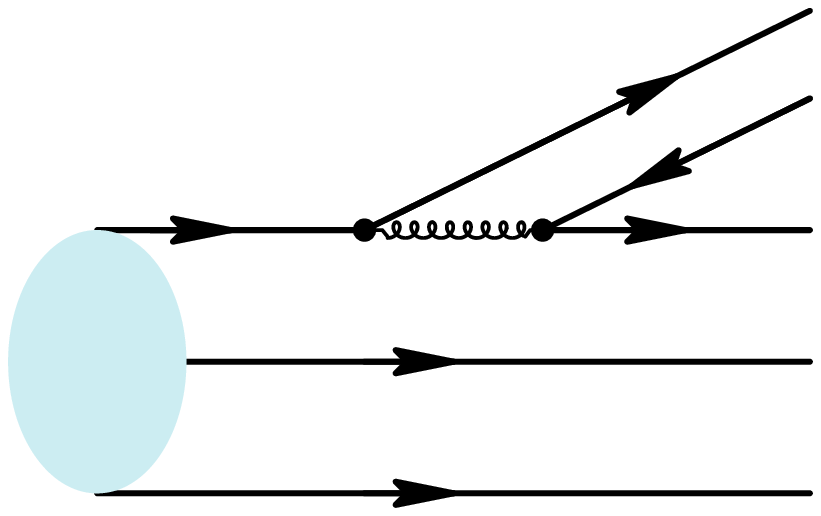}}
\put(33,0){\epsfxsize=55\unitlength \epsfbox{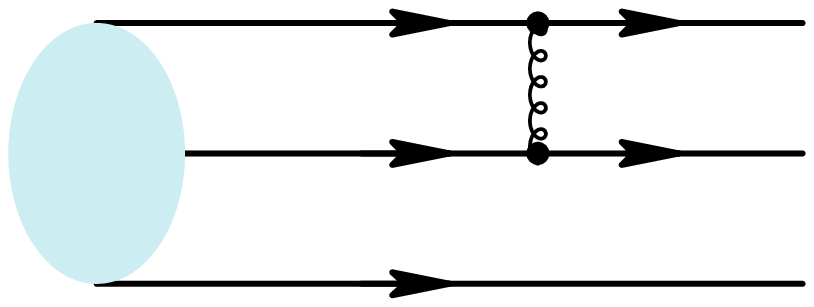}}
\put(85,0){\epsfxsize=55\unitlength \epsfbox{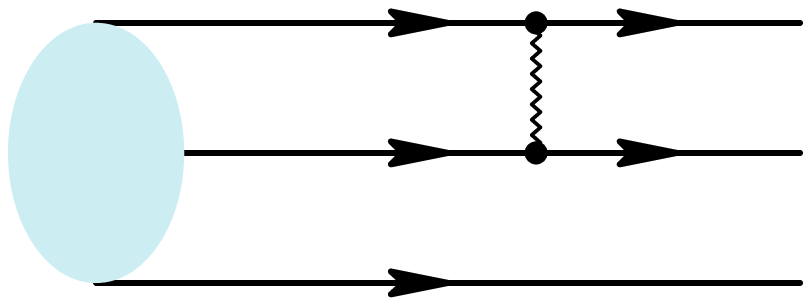}}
\put(-5,7.5){$\Xi^0_{bc}$}
\put(43,7.5){$\Xi^0_{bc}$}
\put(95,7.5){$\Xi^0_{bc}$}
\put(1.5,16.5){$\scriptstyle b$}
\put(12.5,10.5){$\scriptstyle c$}
\put(12.5,3.7){$\scriptstyle d$}
\put(21.5,23){$\scriptstyle s,d$}
\put(23,18.8){$\scriptstyle c$}
\put(26,15.7){$\scriptstyle c$}
\put(60.5,16.5){$\scriptstyle b$}
\put(60.5,10.5){$\scriptstyle c$}
\put(60.5,3.7){$\scriptstyle d$}
\put(71,10.5){$\scriptstyle c$}
\put(70,16.5){$\scriptstyle s,d$}
\put(112.5,16.5){$\scriptstyle c$}
\put(112.5,10.5){$\scriptstyle b$}
\put(112.5,3.7){$\scriptstyle d$}
\put(122.5,16.5){$\scriptstyle s,d$}
\put(123.5,10.5){$\scriptstyle c$}
\end{picture}
        \caption{The penguins and weak scattering diagrams.}
    \label{fig:2}
  \end{center}
\end{figure}

The exclusive modes make the penguin terms to be excluded, since the
penguins add an even number of charmed quarks, i.e. two or zero,
while the final state contains two charmed quarks including one from
the $b$ decay and one from the initial state. By the same reason the
diagrams with the weak scattering of two constituents, i.e. the
charmed and beauty quarks in the $\Xi^0_{bc}$ baryon, are also
excluded for the given final state (see Fig. \ref{fig:2}).

\begin{figure}[th]
  \begin{center}
\setlength{\unitlength}{1mm}
\begin{picture}(90,25)
\put(-15,0){\epsfxsize=55\unitlength \epsfbox{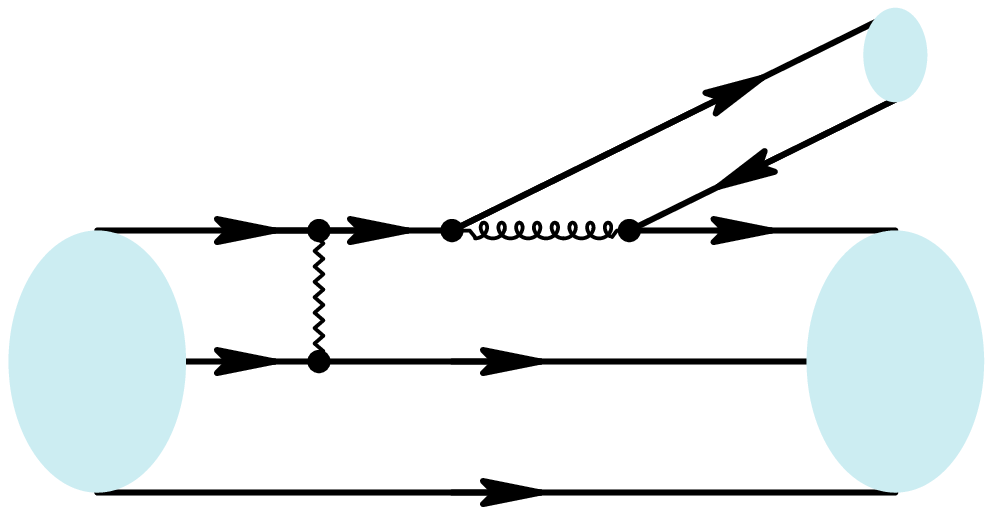}}
\put(35,0){\epsfxsize=55\unitlength \epsfbox{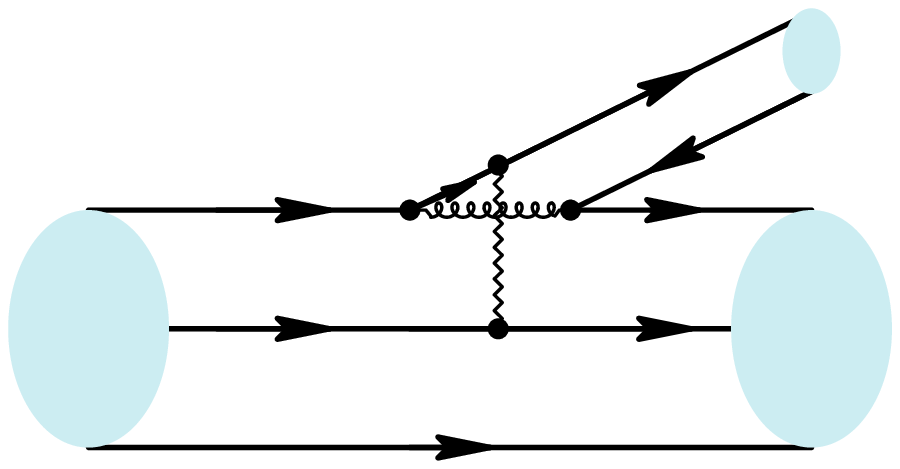}}
\put(-9,7.5){$\Xi^+_{bc}$}
\put(29.5,7.5){$\Xi^+_{c}$}
\put(33,22){$\bar D^0$}
\put(41,7.5){$\Xi^+_{bc}$}
\put(79.5,7.5){$\Xi^+_{c}$}
\put(83,22){$\bar D^0$}
\put(-.5,16.5){$\scriptstyle b$}
\put(6.5,16.5){$\scriptstyle u$}
\put(12.5,10.5){$\scriptstyle s$}
\put(-.5,10.5){$\scriptstyle c$}
\put(12.5,3.7){$\scriptstyle d$}
\put(22.5,23){$\scriptstyle u$}
\put(23,18.8){$\scriptstyle c$}
\put(26,15.7){$\scriptstyle c$}
\put(54.5,16.5){$\scriptstyle b$}
\put(54.5,10.5){$\scriptstyle c$}
\put(62.5,17.7){$\scriptstyle b$}
\put(62.5,3.7){$\scriptstyle d$}
\put(72.5,23){$\scriptstyle u$}
\put(73,18.8){$\scriptstyle c$}
\put(76,15.7){$\scriptstyle c$}
\put(72.,10.5){$\scriptstyle s$}
\end{picture}
    \caption{The diagrams for the weak scattering of $b$ and $c$ quarks
    contributing to the transition $\Xi^0_{bc}\to \bar D^0\Xi^0_c.$}
    \label{fig:2a}
  \end{center}
\end{figure}

The weak scattering of $b$ quark off the charmed quark in the
initial state can contribute in the next order in $\alpha_s$ as
shown in Fig. \ref{fig:2a}. Nevertheless, we see that such the
diagrams have the same weak-interaction structure as at the tree
level. Therefore, they do not break the symmetries under
consideration. The magnitude of $\alpha_s$-correction to the
absolute values of corresponding decay widths is discussed in
Section \ref{sec:2}.

\begin{figure}[th]
  \begin{center}
\setlength{\unitlength}{1mm}
\begin{picture}(80,50)
\put(10,7){\epsfxsize=75\unitlength \epsfbox{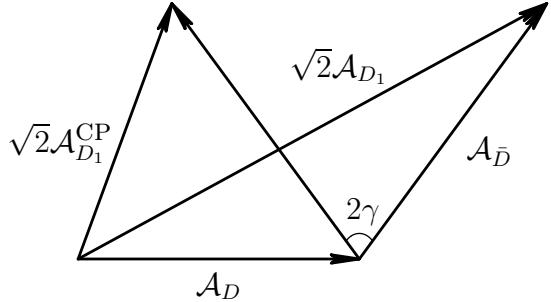}}
\put(32,4){${\cal A}_{D}$} \put(68,22){${\cal A}_{\bar D}$}
\put(52,14){$2\gamma$} \put(44.5,33){$\sqrt{2}{\cal A}_{D_1}$}
\put(7,23){$\sqrt{2}{\cal A}_{D_1}^{\mbox{\sc cp}}$}
\end{picture}
    \caption{The reference-triangles.}
    \label{fig:3}
  \end{center}
\end{figure}

Thus, the CP-odd phases of decays under consideration are determined
by the tree-level diagrams shown in Fig. \ref{fig:1}. Therefore, we
can write down the amplitudes in the following form:
\begin{equation}
  \label{eq:3}
  {\cal A}(\Xi^0_{bc}\to \Xi^0_c\bar D^0) \stackrel{\mbox{\tiny def}}{=}
  {\cal A}_{\bar D} = V_{ub} V^*_{cs}\cdot {\cal M}_{\bar D},\qquad
  {\cal A}(\Xi^0_{bc}\to \Xi^0_c D^0) \stackrel{\mbox{\tiny def}}{=}
  {\cal A}_{D} = V_{cb} V^*_{us}\cdot {\cal M}_{D},
\end{equation}
where ${\cal M}_{\bar D,\,D}$ denote the CP-even factors depending on
the dynamics of strong interactions. Using the definition of angle
$\gamma$
$$
\gamma \stackrel{\mbox{\tiny def}}{=} -{\rm arg}\left[\frac{V_{ub} V^*_{cs}}
{V_{cb} V^*_{us}}\right],
$$
for the CP-conjugated channels\footnote{For the sake of simplicity
  we put the overall phase of arg $V_{cb} V^*_{us}=0$, which
  corresponds to fixing the representation of the CKM matrix, e.g. by
  the Wolfenstein form \cite{Wolf}.} we find
\begin{equation}
  \label{eq:4}
  {\cal A}(\bar \Xi^0_{bc}\to \bar \Xi^0_c D^0)= e^{2{\rm i}\gamma}
  {\cal A}_{\bar D}, \qquad
  {\cal A}(\bar \Xi^0_{bc}\to \bar \Xi^0_c \bar D^0)=  {\cal A}_{D}.
\end{equation}
We see that the corresponding widths for the decays to the flavor
tagged modes coincide with the CP-conjugated ones. However, the story
can be continued by using the definition of CP-eigenstates for the
oscillating $D^0\leftrightarrow \bar D^0$ system\footnote{The
  suppressed effects of CP-violation in the oscillations of neutral
  $D$ mesons are irrelevant here, and we can safely neglect them.},
$$
D_{1,\,2} =\frac{1}{\sqrt{2}}(D^0\pm \bar D^0),
$$
so that we straightforwardly get
\begin{eqnarray}
  \label{eq:5}
 \sqrt{2}{\cal A}(\Xi^0_{bc}\to \bar \Xi^0_c D_1)
 \stackrel{\mbox{\tiny def}}{=}
 \sqrt{2} {\cal A}_{D_1} &=& {\cal A}_{\bar D}+{\cal A}_{D},\\[2mm]
 \sqrt{2}{\cal A}(\bar \Xi^0_{bc}\to \bar \Xi^0_c D_1)
 \stackrel{\mbox{\tiny def}}{=}
 \sqrt{2} {\cal A}_{D_1}^{\mbox{\sc cp}}& = &e^{2{\rm i}\gamma}
 {\cal A}_{\bar D}+{\cal A}_{D}.
\label{eq:6}
\end{eqnarray}
The complex numbers entering (\ref{eq:5}) and (\ref{eq:6}) establish
two triangles with the definite angle $2\gamma$ between the vertex
positions as shown in Fig. \ref{fig:3}.
Thus, due to the unitarity, the measurement of four absolute values
\begin{eqnarray}
  \label{eq:7}
  |{\cal A}_{\bar D}| = |{\cal A}(\Xi^0_{bc}\to \Xi^0_c \bar D^0)|,
&\quad &
  |{\cal A}_{D}| = |{\cal A}(\Xi^0_{bc}\to \Xi^0_c D^0)|, \nonumber\\[2mm]
  |{\cal A}_{D_1}| = |{\cal A}(\Xi^0_{bc}\to \Xi^0_c D_1)|,
&\quad &
  |{\cal A}_{D_1}^{\mbox{\sc cp}}| = |{\cal A}(\bar\Xi^0_{bc}\to \bar \Xi^0_c D_1)|,
\end{eqnarray}
can constructively reproduce the angle $\gamma$ in the model-independent way.

The above triangle-ideology can be implemented for the analogous
decays to the excited states of charmed hyperons in the final state.

The residual theoretical challenge is to evaluate the
characteristic widths or branching fractions. We address this
problem and analyze the color structure of amplitudes. So, we find
that the matrix elements under consideration have the same
magnitude of color suppression ${\cal A}\sim O(1/\sqrt{N_c})$,
while the ratio of relevant CKM-matrix elements,
$$
\left|\frac{V_{ub} V^*_{cs}}{V_{cb} V^*_{us}}\right|\sim O(1)
$$
with respect to the small parameter of Cabibbo angle,
$\lambda = \sin\theta_C$, which one can easily find in the Wolfenstein
parametrization
$$
V_{\mbox{\sc ckm}} = \left(\begin{array}{ccc}
V_{ud} & V_{us} & V_{ub} \\[2mm]
V_{cd} & V_{cs} & V_{cb} \\[2mm]
V_{td} & V_{ts} & V_{tb}
\end{array}\right)
=
\left(\begin{array}{ccc}
1-\frac{1}{2}\lambda^2 & \lambda & A \lambda^3 (\rho-{\rm i}\eta) \\[2mm]
-\lambda & 1-\frac{1}{2}\lambda^2  & A \lambda^2 \\[2mm]
A \lambda^3 (1-\rho-{\rm i}\eta) & -A \lambda^2 & 1
\end{array}\right).
$$
Thus, we expect that the sides of the reference-triangles are of
the same order of magnitude, which makes the method to be a
realistic way for extracting the angle $\gamma$.

In Section \ref{sec:2} we classify the diagrams for the decays of
doubly heavy baryons $\Xi^{0,\,+}_{bc}$ and $\Omega^0_{bc}$ by the
color and weak-interaction structures. Section \ref{sec:3} is devoted
to the numerical estimates in the framework of a potential model. The
results are summarized in Conclusion.

\section{Color structures}
\label{sec:2}

Let us remind a general framework of the $1/N_c$-expansion. Its
physical meaning at $N_c\to \infty$ quite reasonably implies that
the quarks are bound by the gluon string, which is not broken by
the quark-antiquark pair creation. So, the excitations of the
ground states are quasi-stable under the strong interactions
producing the decays described by the $1/N_c$-suppressed terms. In
this method we have got the following scaling rules of color
structures:
\begin{enumerate}
\item
The meson wavefunction
$$
\Psi_M \sim\frac{1}{\sqrt{N_c}}\,{\delta^i}_j.
$$
\item
The baryon wavefunction
$$
\Psi_B \sim\frac{1}{\sqrt{N_c!}}\,\epsilon_{i[1]\ldots i[N_c]}.
$$
\item
The coupling constant
$$
\alpha_s \sim \frac{1}{{N_c}}.
$$
\item
The Casimir operators
$$
C_A=N_c,\qquad C_F=\frac{N_c^2-1}{2N_c}\sim O(N_c).
$$
\item The Fierz relation for the generators of SU($N_c$) group in
the fundamental representation
$$
{t^{A\,i}}_j\, {t^{A\,k}}_m =\frac{C_F}{N_c}\, {\delta^i}_m\,{\delta^k}_j
- \frac{1}{N_c}\, {t^{A\,i}}_m\,{t^{A\,k}}_j.
$$
\item The baryon structure constant
$$
C_B = -\frac{N_c+1}{2N_c}\sim O(1).
$$
It gives the color factor emerging in the connection of two quark lines entering the baryon by the gluon line (see Fig. \ref{fig:4}).
\begin{figure}[th]
  \begin{center}
\setlength{\unitlength}{1mm}
\begin{picture}(80,25)
\put(0,0){\epsfxsize=75\unitlength \epsfbox{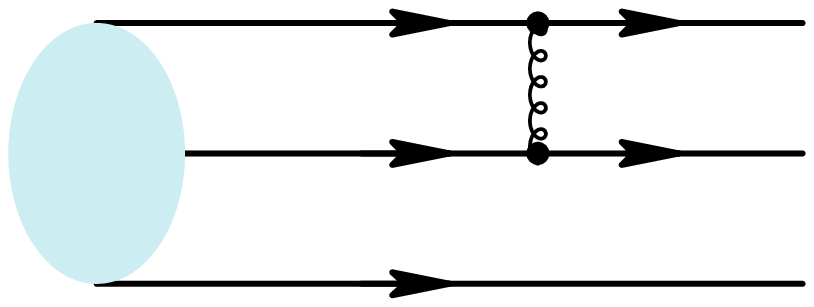}}
\end{picture}
    \caption{The connection of baryon structure constant.}
    \label{fig:4}
  \end{center}
\end{figure}
\end{enumerate}

Next, the non-leptonic weak Lagrangian has a form  typically
given by the following term \cite{BBL}:
\begin{equation}
  \label{eq:7a}
  {\cal H}_{\rm eff} = \frac{G_F}{2\sqrt{2}}\,
V_{cb}(\bar b^i\Gamma_\mu c_j)\,
V^*_{us}(\bar u^k\Gamma^\mu s_l)\,
C_{\pm}\left({\delta^i}_j\,{\delta^k}_l\pm{\delta^i}_l\,{\delta^k}_j\right)
+\ldots
\end{equation}
where $\Gamma_\mu =\gamma_\mu(1-\gamma_5)$, and the Wilson coefficients
$$
C_{\pm}\sim O(1)
$$
in the $1/N_c$-expansion.

Then, we can proceed with the analysis of decays under
consideration.

\subsection{$\boldsymbol \Xi^0_{bc}$}
\label{sec:2.1}

All three diagrams shown in Fig. \ref{fig:1} have the same order
in $1/N_c$, i.e.
$$
{\cal A}_1 \sim {\cal A}_2 \sim {\cal A}_3 \sim  \frac{1}{\sqrt{N_c}}.
$$
More definitely we get the color factors
\begin{equation}
  \label{eq:8}
  {\cal F}^c_1 =\sqrt{N_c}\,a_2, \qquad
  {\cal F}^c_2 =\sqrt{N_c}\,a_2, \qquad
  {\cal F}^c_3 =\frac{C_-}{\sqrt{N_c}} =(a_1-a_2)\,
\frac{\sqrt{N_c}}{N_c-1},
\end{equation}
where
\begin{eqnarray}
  \label{eq:9}
  a_1 &=& \frac{1}{2N_c}\,\left[C_+(N_c+1)+C_-(N_c-1)\right],\\[1mm]
  a_2 &=& \frac{1}{2N_c}\,\left[C_+(N_c+1)-C_-(N_c-1)\right].
\end{eqnarray}
Thus, we have to calculate the three diagrams given above in the
leading order in $1/N_c$.

\subsection{$\boldsymbol \Omega^0_{bc}$}
\label{sec:2.2}
\begin{figure}[th]
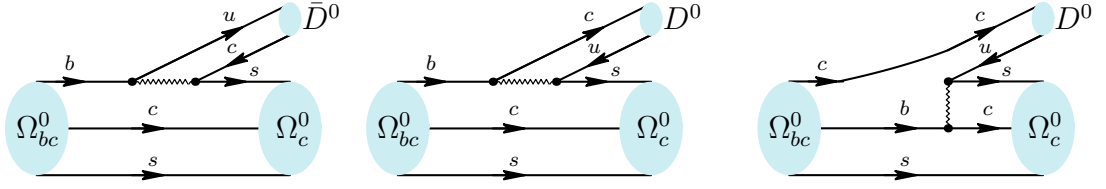

  \begin{center}
\setlength{\unitlength}{1mm}
\begin{picture}(120,25)
\put(-15,0){\epsfxsize=55\unitlength \epsfbox{n1.eps}}
\put(33,0){\epsfxsize=55\unitlength \epsfbox{n1.eps}}
\put(85,0){\epsfxsize=55\unitlength \epsfbox{n2.eps}}
\put(-5,7.5){$\Omega^0_{bc}$}
\put(29.5,7.5){$\Omega^0_{c}$}
\put(33,22){$\bar D^0$}
\put(43,7.5){$\Omega^0_{bc}$}
\put(77.5,7.5){$\Omega^0_{c}$}
\put(81,22){$D^0$}
\put(95,7.5){$\Omega^0_{bc}$}
\put(129.5,7.5){$\Omega^0_{c}$}
\put(133,22){$D^0$}
\put(1.5,16.5){$\scriptstyle b$}
\put(12.5,10.5){$\scriptstyle c$}
\put(12.5,3.7){$\scriptstyle s$}
\put(22.5,23){$\scriptstyle u$}
\put(23,18.8){$\scriptstyle c$}
\put(26,15.7){$\scriptstyle s$}
\put(49.5,16.5){$\scriptstyle b$}
\put(60.5,10.5){$\scriptstyle c$}
\put(60.5,3.7){$\scriptstyle s$}
\put(70.5,23){$\scriptstyle c$}
\put(71,18.8){$\scriptstyle u$}
\put(74,15.7){$\scriptstyle s$}
\put(101.5,16.5){$\scriptstyle c$}
\put(112.5,10.5){$\scriptstyle b$}
\put(112.5,3.7){$\scriptstyle s$}
\put(122.5,23){$\scriptstyle c$}
\put(123,18.8){$\scriptstyle u$}
\put(126,15.7){$\scriptstyle s$}
\put(123.5,10.5){$\scriptstyle c$}
\end{picture}
    \caption{The diagrams of $b$-quark decay contributing to the weak
     transitions $\Xi^0_{bc}\to D^0\Omega^0_c$ and
     $\Xi^0_{bc}\to \bar D^0\Omega^0_c.$}
    \label{fig:5}
  \end{center}
\end{figure}

The decay modes
$$
\Omega^0_{bc}\to D^0\Omega^0_c, \quad\mbox{and}\quad
\Omega^0_{bc}\to \bar D^0\Omega^0_c
$$
are described by the diagrams shown in Fig. \ref{fig:5} similar to
those of Fig. \ref{fig:1}. The only difference is the replacement of
$d$ quark by the strange one, that should be taken into account by
the anti-symmetrization of wavefunction in the final state, i.e. the baryon
structure of $\Omega^0_{c}$, which results in the vector-spin state of the
doubly strange diquark.
Then, the appropriate color factors are given in (\ref{eq:8}).

The diagrams for the decay modes
$$
\Omega^0_{bc}\to D^0\Xi^0_c, \quad\mbox{and}\quad
\Omega^0_{bc}\to \bar D^0\Xi^0_c
$$
can be obtained from those of Fig. \ref{fig:5} by the replacement
of weak currents $u\to s$ by $u\to d$ and $c\to s$ by $c\to d$. The
triangle ideology is effective in the case under consideration too.
The color factors are given by (\ref{eq:8}) again. However, the
amplitude of $\Omega^0_{bc}\to \bar D^0\Xi^0_c$ is suppressed by the
CKM-matrix factor of
$$
\left|\frac{V_{ub}V^*_{cd}}{V_{cb}V^*_{ud}}\right|\sim O(\lambda^2),
$$
which implies that the corresponding side of reference-triangle
will be much less than another. In practice, the relatively large
branching of decay to the $D^0$ meson should be measured with
extremely high accuracy in order to make a sense in the
reconstruction of the triangle with the relatively small side
determined by the branching of decay to $\bar D^0$ meson.

However, with an expected event rate discussed below, in the
nearest future there is no opportunity to observe the $\Omega$
triangles in practice.

\subsection{$\boldsymbol \Xi^+_{bc}$}
\label{sec:2.3}

\begin{figure}[th]
  \begin{center}
\setlength{\unitlength}{1mm}
\begin{picture}(90,25)
\put(-15,0){\epsfxsize=55\unitlength \epsfbox{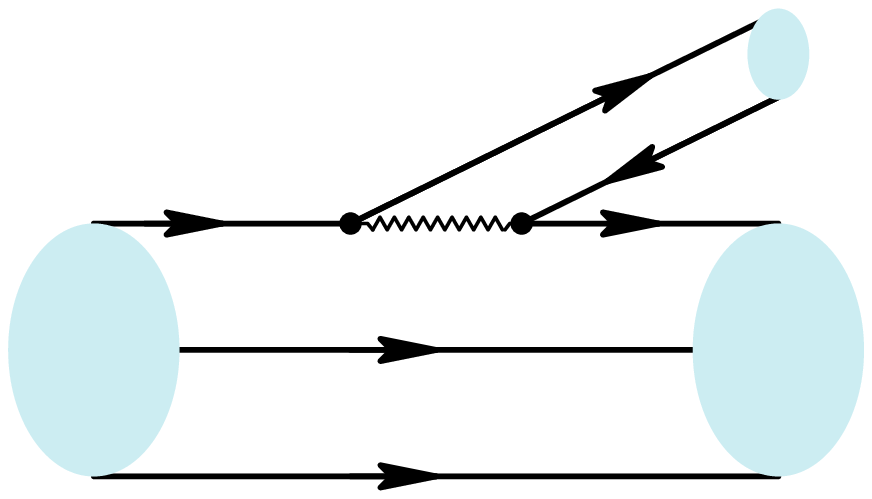}}
\put(35,0){\epsfxsize=55\unitlength \epsfbox{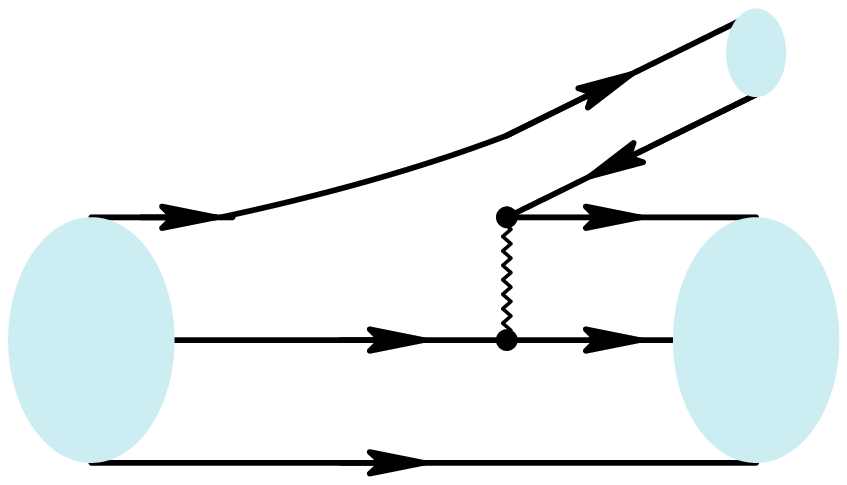}}
\put(-5,7.5){$\Xi^+_{bc}$}
\put(29.5,7.5){$\Xi^+_{c}$}
\put(33,22){$\bar D^0$}
\put(45,7.5){$\Xi^+_{bc}$}
\put(79.5,7.5){$\Xi^+_{c}$}
\put(83,22){$\bar D^0$}
\put(1.5,16.5){$\scriptstyle b$}
\put(12.5,10.5){$\scriptstyle c$}
\put(12.5,3.7){$\scriptstyle u$}
\put(22.5,23){$\scriptstyle u$}
\put(23,18.8){$\scriptstyle c$}
\put(26,15.7){$\scriptstyle s$}
\put(51.5,16.5){$\scriptstyle u$}
\put(62.5,10.5){$\scriptstyle b$}
\put(62.5,3.7){$\scriptstyle c$}
\put(72.5,23){$\scriptstyle u$}
\put(73,18.8){$\scriptstyle c$}
\put(76,15.7){$\scriptstyle s$}
\put(73.5,10.5){$\scriptstyle u$}
\end{picture}
    \caption{The diagrams of $b$-quark decay contributing to the weak
     transition $\Xi^+_{bc}\to \bar D^0\Xi^+_c.$}
    \label{fig:6}
  \end{center}
\end{figure}

The diagrams for the decay
$$
\Xi^+_{bc}\to \bar D^0\Xi^+_c
$$
are shown in Fig. \ref{fig:6}, where the negative relative sign
caused by the Pauli interference should be taken into account.
The corresponding color factors are given by ${\cal F}^c_2$ and
${\cal  F}^c_3$ in (\ref{eq:8}).

\begin{figure}[th]
  \begin{center}
\setlength{\unitlength}{1mm}
\begin{picture}(90,25)
\put(-15,0){\epsfxsize=55\unitlength \epsfbox{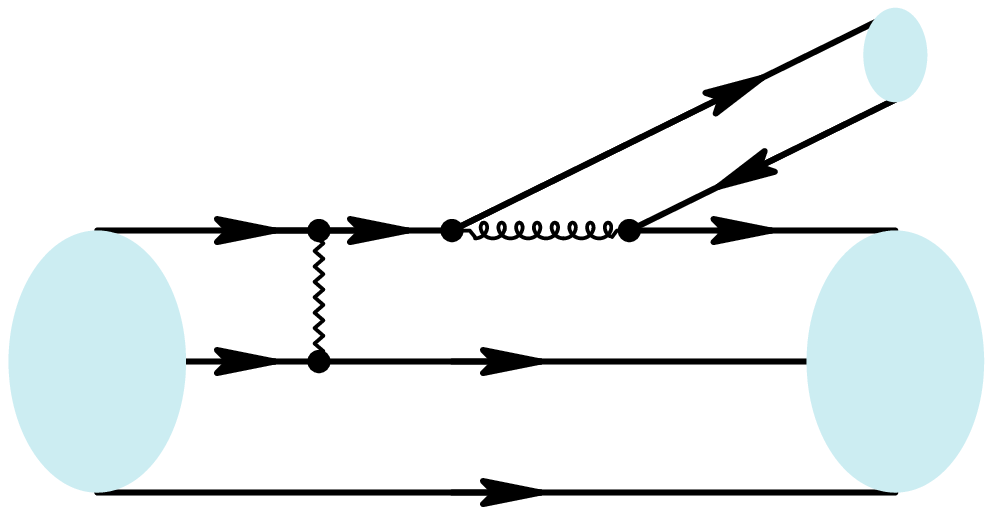}}
\put(35,0){\epsfxsize=55\unitlength \epsfbox{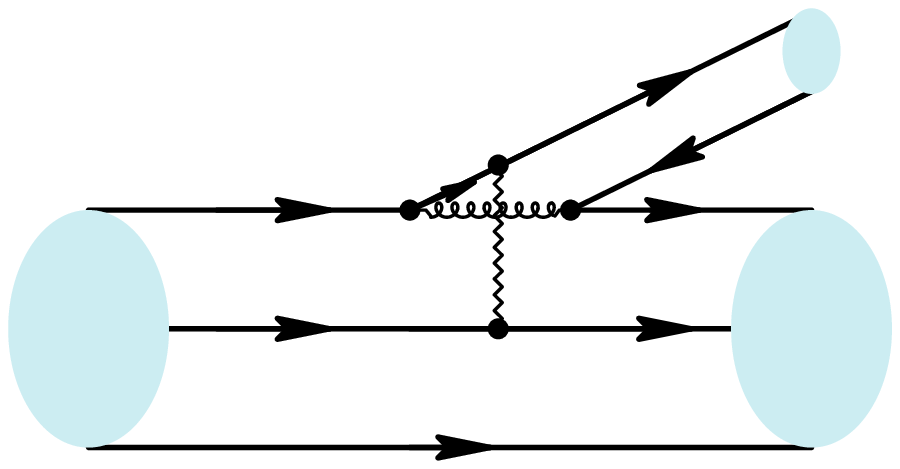}}
\put(-9,7.5){$\Xi^+_{bc}$}
\put(29.5,7.5){$\Xi^+_{c}$}
\put(33,22){$D^0$}
\put(41,7.5){$\Xi^+_{bc}$}
\put(79.5,7.5){$\Xi^+_{c}$}
\put(83,22){$D^0$}
\put(-.5,16.5){$\scriptstyle b$}
\put(6.5,16.5){$\scriptstyle c$}
\put(12.5,10.5){$\scriptstyle s$}
\put(-.5,10.5){$\scriptstyle u$}
\put(12.5,3.7){$\scriptstyle c$}
\put(22.5,23){$\scriptstyle c$}
\put(23,18.8){$\scriptstyle u$}
\put(26,15.7){$\scriptstyle u$}
\put(54.5,16.5){$\scriptstyle b$}
\put(54.5,10.5){$\scriptstyle u$}
\put(62.5,17.7){$\scriptstyle b$}
\put(62.5,3.7){$\scriptstyle c$}
\put(72.5,23){$\scriptstyle c$}
\put(73,18.8){$\scriptstyle u$}
\put(76,15.7){$\scriptstyle u$}
\put(72.,10.5){$\scriptstyle s$}
\end{picture}
    \caption{The diagrams for the weak scattering of $b$ and $u$ quarks
    contributing to the transition $\Xi^+_{bc}\to D^0\Xi^+_c.$}
    \label{fig:7}
  \end{center}
\end{figure}

The consideration of decays into the $D^0$ meson is more complicated
because of the weak scattering of constituent $b$ and $u$ quarks as
shown in Fig. \ref{fig:7}.
We can easily find that the gluon emission from the other quark
lines is suppressed by the color factor since such the exchange by
the gluon leads to the baryon color-structure factor $C_B$ in
contrast to the $C_F$ in the diagrams shown in Fig. \ref{fig:7}.
The spectator charmed quark cannot emit the virtual gluon, since
the quark line should be on mass shell up to the small
virtualities about the relative momentum of the quark inside the
baryon.

\begin{figure}[th]
  \begin{center}
\setlength{\unitlength}{1mm}
\begin{picture}(90,25)
\put(-15,0){\epsfxsize=55\unitlength \epsfbox{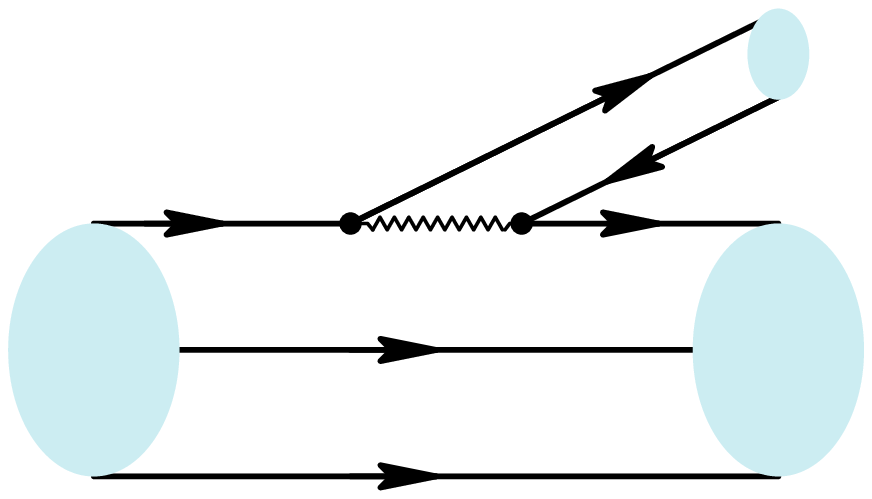}}
\put(35,0){\epsfxsize=55\unitlength \epsfbox{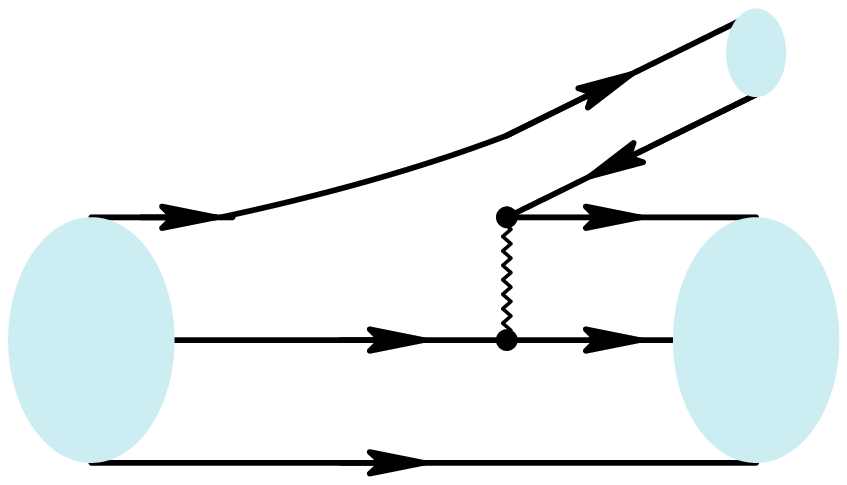}}
\put(-5,7.5){$\Xi^+_{bc}$}
\put(29.5,7.5){$\Xi^+_{c}$}
\put(33,22){$D^0$}
\put(45,7.5){$\Xi^+_{bc}$}
\put(79.5,7.5){$\Xi^+_{c}$}
\put(83,22){$D^0$}
\put(1.5,16.5){$\scriptstyle b$}
\put(12.5,10.5){$\scriptstyle c$}
\put(12.5,3.7){$\scriptstyle u$}
\put(22.5,23){$\scriptstyle c$}
\put(23,18.8){$\scriptstyle u$}
\put(26,15.7){$\scriptstyle s$}
\put(51.5,16.5){$\scriptstyle c$}
\put(62.5,10.5){$\scriptstyle b$}
\put(62.5,3.7){$\scriptstyle u$}
\put(72.5,23){$\scriptstyle c$}
\put(73,18.8){$\scriptstyle u$}
\put(76,15.7){$\scriptstyle s$}
\put(73.5,10.5){$\scriptstyle c$}
\end{picture}
    \caption{The diagrams of $b$-quark decay contributing to the weak
     transition $\Xi^+_{bc}\to D^0\Xi^+_c.$}
    \label{fig:8}
  \end{center}
\end{figure}

Further, the color factors for the diagrams in Fig. \ref{fig:7} have
the form
$$
{\cal F}^c\sim \frac{1}{\sqrt{N_c}}.
$$
Thus, these factors are of the same order of magnitude as for the
decay amplitudes shown in Fig. \ref{fig:8}.

Finally, in this section we have analyzed the color and
weak-interaction structures of decay amplitudes and isolate those
of the largest magnitude, while an illustrative numerical estimate
is presented in the next section.

\section{Numerical estimates}
\label{sec:3}

In this section we formulate the framework of a potential model,
which allows us to evaluate the characteristic widths and
branching ratios for the modes under study.

Let us consider the decay of $\Xi^0_{bc}\to \Xi^0_c\bar D^0$.
So, we define the doubly heavy baryon state in its rest frame by the
following form:
\begin{eqnarray}
|\Xi^0_{bc}  \rangle &=& \sqrt{2M_1}
\int\frac{{\rm d}^3k}{(2\pi)^3}\,\frac{{\rm d}^3q}{(2\pi)^3}\,
\Psi_{bc}({\boldsymbol k})\,\Psi_{d}({\boldsymbol q})\,
\frac{1}{\sqrt{N_c!}}\,\epsilon^{ijk}\,
\left( b^{\rm T}(v_1)\,{\rm C}\,\frac{\gamma_5}{\sqrt{2}}\,c(v_1)\right)
\,d(v_1) \cdot\nonumber\\[1mm]
&&\hspace*{1.7cm}
a^{\dagger}_{i}[b](\boldsymbol k)\,a^{\dagger}_{j}[c](-\boldsymbol k)\,
a^{\dagger}_{k}[d](\boldsymbol q)|0\rangle,
\label{eq:10}
\end{eqnarray}
where $\rm C$ is the charge-conjugation matrix, and we have introduced
the following notations:
\begin{itemize}
\item the relativistic normalization-factor $\sqrt{2M_1}$ with $M_1$
  being the baryon mass in the initial state,
\item the wavefunction $\Psi$ given by two factors of the doubly heavy
  diquark and the $d$ quark,
\item the color wavefunction determined by $\epsilon^{ijk}/\sqrt{N_c!}$,
\item the spin wavefunction of scalar doubly heavy diquark as
  determined by the spinor factor $\gamma_5/\sqrt{2}$,
\item the quark spinors depending on the four-velocity of the baryon,
  $v_1$, as normalized by the condition of sum over the polarization
  states, say, $$\sum b(v_1)\bar b(v_1) = \frac{1}{2}(1+\slashchar{v}_1),$$
\item $a^\dagger$ denoting the creation operator marked by the flavor,
  color and momentum in an appropriate way.
\end{itemize}
In (\ref{eq:10}), we use the momentum-space wavefunction in the form of
$$
\Psi(k^2)= \left(\frac{8\pi}{\omega^2}\right)^{3/4}\,
\exp\left[\frac{k^2}{\omega^2}\right],
$$
where the four-vector satisfies the condition
$$
k\cdot v_1 =0.
$$
The model parameter $\omega$ is related with the wavefunction
$\tilde \Psi(0)$ at the origin in the configuration space. Such the
quantities were calculated by solving the Schr\"odinger equation with
the static potential as described in \cite{KKO}, so that
$$
\tilde\Psi_{bc}(0) = 0.73\;{\rm GeV}^{3/2},\qquad
\tilde\Psi_{d}(0) = 0.53\;{\rm GeV}^{3/2},
$$
while for the $\Xi^0_{c}$ baryon we take
$$
\tilde\Psi_{cs}(0) = 0.61\;{\rm GeV}^{3/2},\qquad
\tilde\Psi_{d}(0) = 0.53\;{\rm GeV}^{3/2}.
$$
We will see that for the mode under consideration the absolute
values of above parameters are not so critical, while their ratios are
important.

Then the matrix element is equal to
\begin{equation}
  \label{eq:11}
  {\cal A}_{\bar D} = \frac{G_F}{\sqrt{2}}\, V^*_{ub}V_{cs}\,
\sqrt{M_1M_2M_D}\,\tilde\Psi_D(0)\,\sqrt{N_c}a_2\cdot {\cal T}
\cdot{\cal O}\cdot
\bar u(v_2)u(v_1),
\end{equation}
where the mass and wavefunction factors as well as the spinors and
four-velocities are transparently denoted, while $\cal T$ is given by the
spin structure of the matrix element
$$
{\cal T} = \frac{1}{8}{\rm Tr}\left[(1+\slashchar{v}_1)(1+\slashchar{v}_2)
\gamma_\mu(1-\gamma_5)\gamma_5(1+\slashchar{v}_D)\gamma^\mu
(1-\gamma_5)
\right],
$$
and $\cal O$ represents the overlapping of wavefunctions, which can
be calculated in the model specified. So, it is presented by the
product of $\xi$ factors
$$
{\cal O} = \xi_{bc\to cs}(y)\cdot \xi_d(y),
$$
where the diquark transition is determined by
\begin{equation}
\xi_{bc\to cs} = \frac{2 \omega_{bc}\omega_{cs}}{\omega_{bc}^2+\omega_{cs}^2}\,
\sqrt{\frac{2\omega_{bc}\omega_{cs}}{\omega_{cs}^2+y^2\omega_{bc}^2}}\,
\exp\left[-\frac{\tilde m^2(y^2-1)}{\omega_{cs}^2+y^2\omega_{bc}^2}\right],
\label{eq:13}
\end{equation}
with $y= v_1\cdot v_2$ and
$$
\tilde m = m_c\,\frac{M_1+m_s-m_b}{m_c+m_s}\approx m_c+
O\left(m_{s,\,d}/m_c\right).
$$
In the given kinematics the product of four-velocities is fixed by
$$
y = \frac{M_1^2+M_2^2-M_D^2}{2M_1M_2}.
$$
The factor of $\xi_d(y)$ has the form similar to (\ref{eq:13}) under
the appropriate substitutions for the values of $\omega$ as well as for
$\tilde m\to m_d$ \cite{Kiselev}.

Having written down eq.(\ref{eq:11}), we have neglected the
$\alpha_s$-corrections following from the diagrams shown in Fig.
\ref{fig:2a}. This approximation is theoretically sound. Indeed, the
gluon virtuality is determined by the expression
$$
k_g^2 = m_c^2 (v_2+v_D)^2 =
m_c^2\,\frac{M_1^2-(M_2-M_D)^2}{2M_2M_D} \approx
{(m_b+m_c)^2}+O(m_{s,\,d}/m_{c,\,b})\gg \Lambda^2_{\mbox{\sc
qcd}}.
$$
The suppression factor of appropriate diagrams is given by
$$
{\cal S}\sim |\tilde \Psi(0)|^2\,\frac{\alpha_s(k_g^2)}{k_g^2}\,
\frac{N_c}{\Delta E_Q},
$$
where
$$
|\tilde\Psi(0)|^2 \sim \Lambda^3_{\mbox{\sc qcd}}
$$
is the characteristic value of wavefunction, and the virtuality of
heavy quark line connected to the virtual gluon is of the order of
$$
\Delta E_Q \sim m_{c,\,b}.
$$
Therefore,
$$
{\cal S} \sim \frac{\Lambda^3_{\mbox{\sc qcd}}}{m_{c,\,b}^3}\,
\frac{1}{\ln m_{c,\,b}/\Lambda_{\mbox{\sc qcd}}}\ll 1.
$$
Similar relations can be found, if we consider the value of
$\alpha_s$-correction due to the weak scattering shown in Fig.
\ref{fig:7}. One can easily find that the corresponding suppression is
given by
$$
\tilde{\cal S} \sim \frac{\Lambda_{\mbox{\sc qcd}}}{m_{c,\,b}}\ll
1,
$$
where we have suggested the constituent mass of light quark
$$
m_{u,\,d}\sim \Lambda_{\mbox{\sc qcd}}.
$$
Therefore, the above $\alpha_s$-corrections can be neglected to the
leading order in the $1/m_Q$-expansion.

Further,
$$
{\cal T} = 2(1+y)\,\frac{M_1-M_2}{M_D},
$$
and we can get the matrix element squared
\begin{equation}
  \label{eq:14}
  \left|{\cal A}_{\bar D}\right|^2 = \frac{G_F^2}{2}\,
\left|V^2_{ub}V^2_{cs}\right|\,M_1M_2(M_1-M_2)^2(1+y)^3\,
f^2_D\cdot {\cal O}^2\cdot (N_c a_2)^2,
\end{equation}
where we have expressed the wavefunction of $D$ meson in terms of its
effective leptonic constant $f_D$,
$$
f_D = 2\sqrt{\frac{3}{M_D}}\,\tilde\Psi(0).
$$
Then, the width is given by
$$
\Gamma[\Xi^0_{bc}\to \Xi^0_c\bar D^0] = \frac{|\boldsymbol k_D|}{16\pi M_1^2}\,
\left|{\cal A}_{\bar D}\right|^2,
$$
where $\boldsymbol k_D$ is the momentum of D meson in the
$\Xi^0_{bc}$ rest-frame.

Numerically, at $M_1\approx 6.9$ GeV \cite{qqM,PMQQq} we find
\begin{equation}
  \label{eq:15}
  \Gamma[\Xi^0_{bc}\to \Xi^0_c\bar D^0] \approx 1.3\cdot 10^{-6}\;{\rm ps}^{-1}
\times
\left|\frac{V^2_{ub}}{0.003^2}\,\frac{V^2_{cs}}{0.975^2}\right|\,
\frac{f^2_D}{(0.222\;{\rm GeV})^2}\cdot {\cal O}^2\cdot \frac{(N_c a_2)^2}{1}.
\end{equation}
Putting \cite{revqq,KOtau}
$$
\tau[\Xi^0_{bc}] =0.27\;{\rm ps},
$$
we get
\begin{equation}
  \label{eq:16}
  {\cal B}[\Xi^0_{bc}\to \Xi^0_c\bar D^0] \approx
3.6\cdot 10^{-6}
\times
\left|\frac{V^2_{ub}}{0.003^2}\,\frac{V^2_{cs}}{0.975^2}\right|\,
\frac{f^2_D}{(0.22\;{\rm GeV})^2}\cdot {\cal O}^2\cdot \frac{(N_c a_2)^2}{1}.
\end{equation}
The calculation of overlap between the wavefunctions is
model-dependent, though one can expect that ${\cal O}\sim 1$. In the
framework of potential model described we get
$$
{\cal O}\approx 0.44,
$$
which gives our final estimate
\begin{equation}
  \label{eq:17}
  {\cal B}[\Xi^0_{bc}\to \Xi^0_c\bar D^0] \approx
0.7\cdot 10^{-6}
\times
\left|\frac{V^2_{ub}}{0.003^2}\,\frac{V^2_{cs}}{0.975^2}\right|\,
\frac{f^2_D}{(0.22\;{\rm GeV})^2}\cdot \frac{{\cal O}^2}{0.2}\cdot
\frac{(N_c a_2)^2}{1}.
\end{equation}
Therefore, we could expect that the other branching ratios are of the
same order of magnitude.

\section{Conclusion}
\label{sec:con}

In this work we have extended the reference-triangle ideology to the
model-independent extraction of CKM-matrix angle $\gamma$ from the set
of branching ratios of doubly heavy baryons exclusively decaying to
the neutral $D$ mesons. Tagging the flavor and CP-eigenstates of such
the $D$ mesons allows one to avoid the uncertainties caused by the QCD
dynamics of quarks.

We have estimated the characteristic branching ratios in the framework
of a potential model, which yields, for example,
$$
{\cal B}[\Xi^0_{bc}\to \Xi^0_c\bar D^0] \approx
0.7\cdot 10^{-6}.
$$
Accepting the above value, we can estimate the rate of events at
the LHC collider in the experiment LHCB or in the BTeV facility at
FNAL. So, the production cross section yields the characteristic
value of about $10^9$ doubly charmed baryons per year
\cite{BRunII}. Next, the estimated branching rations for the
charmed strange baryons are measured for $\Xi_c^+$, so that the
detection of charged particles in the final state would cover
about 50\% of decay events with $\Xi_c^+$, which we accept for the
optimistic estimates. The efficiency of observing the neutral
charmed meson crucially depends on the possibility for detecting
the neutral kaons and pions. So, removing the neutral kaons and
pions could kill the opportunity offered in the present paper. We
consider the optimistic case with the required detection at work,
which gives the efficiency for observing the $D^0$ decay at the
level of 25\%. Therefore, putting the vertex reconstruction
efficiency equal to 10\%, we can expect the observation  of about
$10^9\cdot 0.7\cdot 10^{-6}\cdot 5\cdot 10^{-1}\cdot 2.5\cdot
10^{-1}\cdot 10^{-1} \approx 11$ events per year. However, the
situation is less optimistic for the detection of CP-eigenstates
of $D^0$. Indeed, the CP-even mode with the $\pi^+\pi^-$ or
$K^+K^-$ final states covers only a $5\cdot 10^{-3}$ fraction of
$D^0$ decays, which makes the observation unreachable. The CP-odd
events with $K_S\pi^0$ can be detected with the branching fraction
about $2.5\cdot 10^{-2}$, which downs the rate to 1 event per
year. Thus, the reconstruction of reference triangle after 10
years of data taking would give, at least, the 30\% accuracy for
the characteristic triangle side, which makes the extraction of
angle $\gamma$ rather academic exercise in the method offered.

The authors thank the organizing Committee of the Heavy Quarkonium
Workshop held in CERN, Nov. 8-11, 2002, and personally to Antonio
Vairo and Nora Brambilla for the invitation and a kind
hospitality. V.V.K. also thanks prof. R.Dzhelyadin for the
possibility to visit CERN in collaboration with the LHCB group, to
which members he expresses a gratitude for a hospitality. We thank
prof. A.K.Likhoded, Drs. Yu.Gouz and V.Romanovsky for useful
discussions and valuable remarks.

This work is in part supported by the Russian Foundation for Basic
Research, grant 01-02-99315.


\end{document}